\documentclass[10pt,conference]{IEEEtran}
\usepackage{epsfig}
\usepackage{url}
\usepackage{subfigure}

\providecommand{\eat}[1]{}

\usepackage{epsfig}
\usepackage{url}
\begin{document}
\title{\huge\bfseries {Investigating Randomly Generated Adjacency Matrices For Their Use In Modeling Wireless Topologies}}
\author{Gautam Bhanage and Sanjit Kaul\\
\{gautamb, sanjit\}@winlab.rutgers.edu \\
WINLAB, Rutgers University, North Brunswick, NJ 08902, USA. }


\maketitle
\thispagestyle{empty}
\let\VERBATIM\verbatim
\def\verbatim{%
\def\verbatim@font{\small\ttfamily}%
\VERBATIM}

\begin{abstract}
Generation of realistic topologies plays an important role in
determining the accuracy and validity of simulation studies. This
study presents a discussion to justify why, and how often randomly
generated adjacency matrices may not not conform to wireless
topologies in the physical world. Specifically, it shows through
analysis and random trials that, \emph{more than $90\%$ of times, a
randomly generated adjacency matrix will not conform to a valid
wireless topology, when it has more than $3$ nodes}. By showing that
node triplets in the adjacency graph need to adhere to rules of a
geometric vector space, the study shows that the number of randomly
chosen node triplets failing consistency checks grow at the order of
$O(base^{3})$, where $base$ is the granularity of the distance
metric. Further, the study models and presents a probability
estimate with which any randomly generated adjacency matrix would
fail realization. This information could be used to design simpler
algorithms for generating \emph{k-connected} wireless topologies.
\end{abstract}
\section{Introduction}
\label{sec:intro}

Simulation studies can be easily setup for wired networks by
generating a random adjacency matrix for modeling a random topology.
As long as finite non-negative entries are chosen for the adjacency
matrix, it could be used to represent a valid wired topology.
However, in this paper we discuss how, and why this may not hold
true in the case of wireless topologies.

Specifically, this study addresses the following questions:
\begin{enumerate}
\item{\emph{Correctness:} } Are randomly generated topologies always valid,
if not under what conditions.
\item{\emph{Frequency of Failure:} } What percentage of randomly generated
matrices are invalid?
\item{\emph{Dominant Failure Factor:} } What feature of the matrix decides
the probability of the topology being invalid?
\item{\emph{Implication: }}Using this understanding,
we propose designing algorithms with a relatively direct approach
for generating k-connected graphs.
\end{enumerate}

Rest of the paper is organized as follows.
Section~\ref{sec:overview_problem_solution} shows an example where a
randomly generated adjacency matrix does not represent a wireless
topology. Section~\ref{sec:modeling} describes the problem
statement, and present our approach for determination of valid
matrices. Section~\ref{sec:random_tests} presents a comparison of
results from random trials with an approximation generated by our
probability function. Finally, we present a brief conclusion.


\section{Discussion}
\label{sec:overview_problem_solution}

\begin{figure}[t]
\vspace{0.25cm}
\begin{center}
\epsfig{figure=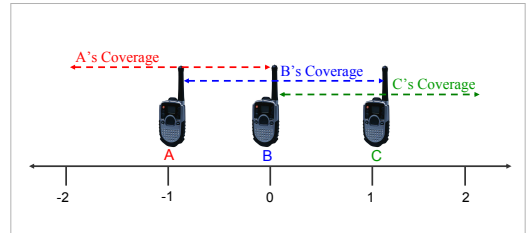,width=2.7in}
\end{center}
\caption{Mapping problem definition as seen in $\Re^1$ space. If we
have node B connected to A and C, we cannot have another node D with
connectivity to A and C but not connected to B on the number line in
$\Re^1$ space.} \label{fig:problem_defn_1d}
\end{figure}

\subsection{Example Of An Invalid Wireless Topology}
We first address the question of whether a randomly generated
adjacency matrix can result in a non-realizable wireless topology.
Figure~\ref{fig:problem_defn_1d} shows the positions of three
previously mapped points A, B, and C in a one dimensional metric
space ($\Re^1$). For this problem, we consider that all nodes have
similar radio capabilities and can communicate with each other only
if they are within \emph{unit} distance of each other. As per this
condition, we have node B connected to nodes A and C.


Now consider a case where the adjacency matrix generating the
topology in Figure~\ref{fig:problem_defn_1d} has an additional entry
for a fourth node D, which has links to A and C but is not within
coverage of node B. Such a wireless topology is physically not
possible in the one dimensional metric space ($\Re^1$). Note that
this problem cannot be solved by using a different channel, since
all the nodes will need to be on the same frequency to be
connected\footnote[1]{We refer to a \emph{connection} between any
nodes 1 and 2, as a general term to signify that 1 and 2 have a
significant SNR to communicate with each other. This
\emph{connectivity} is at the layer-1 and is independent of any
access control mechanisms used at a higher layer in the network
stack.}. It is also important to observe that this failure occurs
even when we do not have any planarity constraints like requiring
non-intersecting graph edges. Non-uniform radio coverage for nodes D
and B also fails to solve the problem. This is because both nodes D
and B need to be on the line, and a non-uniform radio coverage in
either directions (left or right) will result in disconnection from
nodes A and/or C.

This problem can be extended to all higher dimensions in metric
spaces $\Re^n$, $n>0$, which could result in invalid physical
topologies. The only factor that varies across these dimensions is
the nature of the wireless coverage. In $\Re^1$, we consider a line
of unit (manhattan) distance on each side of the node, in $\Re^2$,
the coverage can be assumed in the form of a unit circle in the
plane (euclidean distance), similarly, a unit sphere in $\Re^3$ and
so on.


\subsection{Problem Statement} Now that we have shown an example of
an invalid wireless topology generation, we will explicitly define
the problem. Consider a network graph G which is generated by a
random adjacency matrix $A_{adj}[\texttt{ }]_{n \times n}$, where
$n$ denotes the number of nodes in the graph G. The individual
entries in $A_{adj}$ will denote the link conditions between
corresponding wireless nodes. In this study, given a specific
$A_{adj}[\texttt{ }]$, we will define a function \emph{F}() to tell
us whether the given adjacency matrix is capable of realizing a
valid wireless topology or not:
\begin{equation}
F: G(A_{adj}[\texttt{ }]_{n \times n}) \mapsto \{Valid, Invalid\}
\end{equation}
Once determined, \emph{F}() can be used as a test for incrementally
adding neighbor nodes to an adjacency matrix for generating
\emph{k-connected} graphs. We also calculate the probability
$(P_{F})$ with which \emph{F}() will fail, which could be used as a
metric for determining the  average number of trials that would be
required for valid wireless topology generation.


\section{Modeling}
\label{sec:modeling}


\subsection{Wireless Topologies $\&$ Vector Spaces}
To determine \emph{F}() defined above, we briefly discuss why the
random adjacency matrix used for Figure~\ref{fig:problem_defn_1d}
fails. If we consider, the first three nodes A, B, C, we observe
that they satisfy the triangle inequality requirements in the
$\Re^{1}$ metric space. Let $\parallel.\parallel$ represent an
arbitrary distance norm. Triangle inequality requirement states that
the sum of the lengths of any two sides (say $\parallel x
\parallel + \parallel y
\parallel_{}$)  has to be greater than the third side ($\parallel
x+y \parallel$). While mapping the fourth node D, with the
requirement $A_{adj}(A,D)=1$, $A_{adj}(D,C)=1$, and
$A_{adj}(B,D)=0$, we observe that the triangle inequality fails for
the node sets $\{B, D, A\}$ and $\{B, D, C\}$. Thus using simple
triangle inequality as a test for the function $F$, i.e by
determining if the generated wireless topology fits in a geometric
vector space, we can classify random matrices as representing valid
or invalid wireless topologies.


\subsection{Estimating Adjacency Matrix Failure Probability $(P_F)$}

A randomly generated adjacency matrix $A_{adj}$ for a wireless
topology with $n$ nodes will fail when any one combination of three
links fails the triangle inequality check. Hence, the probability of
at least one failure is:
$P_{F} = 1 - P_{NF},$
where the $P_{NF}$ is the probability that no combination in the
adjacency graphs fails the triangle inequality check. Thus, $P_{F}$
can be calculated as:
\begin{equation}
P_{F} = 1 - (1 - P_{\triangle} )^{N_{pairs}},
\end{equation}
where $N_{pairs}$ denotes the number of combinations of nodes
checked in a randomly generated matrix, and $P_{\triangle}$ denotes
the probability of failure of the triangle inequality on any random
adjacency triplet. We define an \emph{adjacency triplet} as any
single combination of three values $A_{adj}(P,Q)$, $A_{adj}(Q,R)$
and $A_{adj}(P,R)$ that describe link conditions between any three
nodes P, Q and R. The $N_{pairs}$ are determined by the number of
non-diagonal entries in the adjacency matrix. Since the adjacency
matrix is representing a wireless topology, it should be symmetric
akin to a metric space distance matrix. Hence, $N_{pairs} =
(\frac{n^2 - n}{2}) \times (\frac{n^2 - n}{2} - 1)$.



\begin{figure}[t]
\begin{center}
\epsfig{figure=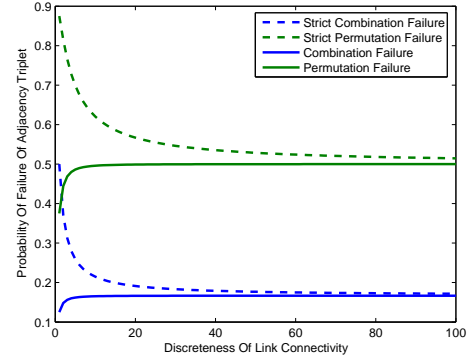,width=2.7in}
\end{center}
\caption{Probability of failure of triangle inequality tests for
unique combinations and permutations of side triplets.}
\label{fig:failures_numberbase}
\end{figure}

\subsection{Determining $P_{\triangle}$}
To estimate $P_{\triangle}$, we use the complete set of adjacency
triplets ($S_3$) described as:
\begin{equation}
\texttt{ }S_3 = \{A_{adj}(P,Q),\texttt{ }A_{adj}(Q,R),\texttt{ }
A_{adj}(P,R)\},
\end{equation}
defined $\forall P,Q,R \in A_{adj}$. To determine $P_{\triangle}$ we
can either use combinations or permutations on $S_3$ to determine
failure probability of combinations. For all such possible
permutations and combinations over $S_3$, we determine
$P_{\triangle}$ by calculating the fraction of adjacency triplets
that fail strict ($\le$) and non-strict ($\leq$) triangle inequality
checks. While evaluating, we vary the \emph{base}, which denotes the
maximum number of discretized values that can be used to represent
the link between two points. E.g. when we chose the base as 1, the
link represented in the adjacency matrix can either take the values
as 0 (off) or 1(on). If the base is 2, the possible values are 0,1,2
and so on. Results for this model are as described in
Figure~\ref{fig:failures_numberbase}.

We observe that the fraction of permutations or combinations
resulting in the triangle inequality failure remain fairly constant,
irrespective of the increasing number base. Also, we note that the
number of combinations being evaluated are growing as $O(base^3)$.
Hence, we conclude that, the number of unique combinations failing
are also increasing at $O(base^3)$ to keep the ratio constant.





\section{Monte Carlo Tests}
\label{sec:random_tests}


In this section we estimate and compare the probability with which a
randomly generated adjacency matrix will fail when mapped as a
wireless topology. We compare our failure probability estimate
($P_{F}$) with the failure probability obtained through randomized
trials. In these comparisons, we use the discreteness of link
qualities (or the discreteness of distance) and the size of wireless
topologies as the two varying parameters.

\subsection{Discreteness Of Link Connectivity Representation}


\begin{figure}[t]
\begin{center}
\epsfig{figure=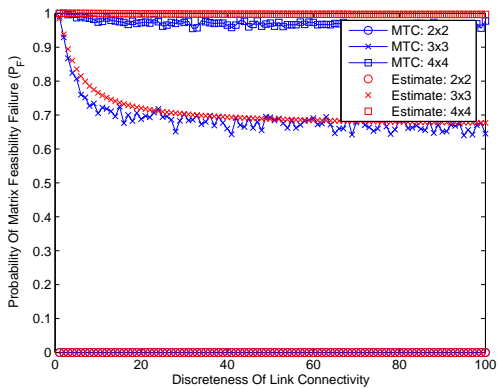,width=2.9in}
\end{center}
\caption{Probability of failure of a randomly generated adjacency
matrix in representing a wireless topology as a function of the
discreteness of distance or connectivity.}
\label{fig:plotx_matrixbase}
\end{figure}

In this test, as with the estimation of $P_{\triangle}$, we vary the
maximum number of discretized values (\emph{base}) that can be used
to represent the link between two points. Results from our estimates
and those from monte-carlo tests are correspondingly marked as
\emph{Estimate:*} and \emph{MTC:*} in the
Figure~\ref{fig:plotx_matrixbase}. The results show that our
estimate of $P_{F}$ is able to closely match the failure probability
obtained from trials of $1000$ randomly generated adjacency matrices
for every distance \emph{base}. For all topology sizes: $2,3,4$
(nodes each), we observe that the probability of the matrix failing
to conform to a wireless topology ($P_{F}$) can be high when the
link connectivity is coarsely described (E.g. on or off). This
result is a direct consequence of: $P_{F} \propto P_{\triangle}$.
Hence, we observe that as $P_{\triangle}$ stabilizes for higher
values of the distance \emph{base}, $P_{F}$ stabilizes too.

%

\begin{figure}[t]
\begin{center}
\epsfig{figure=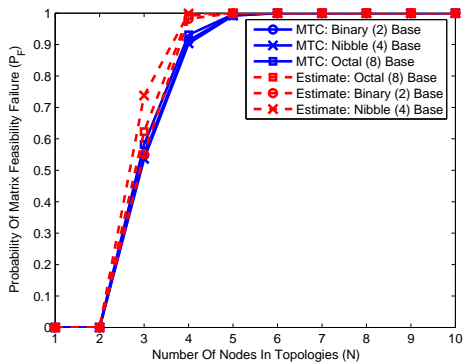,width=2.7in}
\end{center}
\caption{Comparison of estimated and observed failure probability as
a function of varying topology size.} \label{fig:plotx_matrixdim}
\end{figure}


\subsection{Impact Of Varying Topology Size}
The size of a topology can be changed by varying the number of
nodes. Edges are not explicitly used as a factor for changing
topology size since the number and type of edges are randomly
decided. In this experiment, we vary the size of the wireless
topology from 1 to 10 nodes. For every topology, an edge can have a
value uniformly distributed among the number of discretized distance
values given by the \emph{base}. We generate $1000$ random matrices
for each topology size.

As shown in the results in Figure~\ref{fig:plotx_matrixdim} the
estimated failure probability $P_{F}$ (denoted by \emph{Estimate:*})
closely matches that obtained from random trials (\emph{MTC:*}).
Further, we observe that failure probability quickly approaches 1.
This matches with our estimate since, $P_{F} \propto N_{pairs}$, and
$N_{pairs}$ increase at least as $O(n^{2})$. An important
implication of this result is that as the size of the wireless
topology goes beyond $3$ nodes, it is almost certain that a randomly
generated adjacency matrix will not conform to a wireless topology.

\section{Related Work}
\label{sec:related}

A class of studies has focussed on enumerating the characteristics
of wired networks~\cite{Doar96abetter,Zegura96howto} that need to be
taken into consideration while generating topologies from random
graphs. Consequently, a parallel area of research is focussed on
efficient generation~\cite{Jurij05Realistic} and improvement of the
features of random graphs to model real wired
networks~\cite{Rodionov04Random}.

With concerns to wireless networks, ~\cite{Bettstetter02On}
investigates the impact of spatial distribution of nodes on the
minimum node degree, and the k-connectivity in random network
graphs. We take a completely opposite view of the problem in
determining if a randomly generated adjacency could be used for
faithfully representing a realistic wireless topology.



\section{Conclusions}
\label{sec:conclusion}
This study describes an approach for determining if randomly
generated adjacency matrices can conform to wireless topologies in
the physical world. It is shown that these random matrices are prone
to failure, specially for topologies with more than 3 nodes. Using
this information, an alternative approach can be taken for random
wireless topology creation. Instead of designing simulation studies
based on random placement of nodes and then generating k-connected
graphs, algorithms could be designed that would iteratively add
nodes to the adjacency graph based on k-connectivity requirements as
long as they do not violate constraints of the geometric space.

\bibliographystyle{IEEEtran} 
\bibliography{ref}

\end{document}